# A Fermionic Background Field Formulation of QCD

**By Dr. Scott Chapman**
scottc@cal.berkeley.edu

**Abstract**
  A new background field formulation of QCD is presented in which the background gluon field is not a classical field, but an operator made up of quantized quark fields. This background field allows colorless quark states to form exact quantum solutions of the QCD equations of motion for any value of the coupling constant. When matrix elements of the Hamiltonian are calculated in the context of these solution states, quark fields and gluon fluctuations completely decouple. Due to decoupling, gluon fluctuations around the background field can be ignored and only the quark part of the Hamiltonian need be considered when comparing colorless quark states. Despite nonlinear terms involving the background field, this pure quark Hamiltonian is completely diagonalizable and leads to energies for colorless quark configurations that are infinitely more negative than those for colored quark configurations. In this way, the method provides an explanation for confinement and can be used to calculate hadron masses directly from QCD.

# Introduction

In recent years, much work has been done in attempting to analytically derive quark confinement mechanisms from QCD. One theory is that the QCD vacuum consists of a gluon condensate [1,2] that causes the vacuum to have a negative energy density. This negative energy density exerts a confining pressure on quarks that balances the outward pressure of quark kinetic energy, leading to stable hadron radii. The MIT bag model presented in [3] was an early phenomenological attempt to model hadrons using this mechanism. Since then many improvements to the bag model have been made, but they still rely on phenomenological interactions and parameters that are not directly derived from QCD.

If a gluon condensate does exist, clearly the simplest approach would be to assume that the vacuum has some constant gluon electric or magnetic field that produces the negative energy density. One could then use mean field theory to make the connection from QCD to the phenomenological models. This approach was taken in [4,5], where it was shown that a mean chromomagnetic gluon field could produce the needed negative energy density. The main problem with this mean field approach is that since the gluon field is a vector field, the existence of a classical mean field implies that the vacuum has some preferred direction in spacetime.

Many approaches have been suggested for getting around this Lorentz invariance problem. Most of these approaches utilize gluon field configurations that couple Lorentz transformations with gauge transformations in such a way that Lorentz invariance is maintained despite the vector nature of the vacuum gluon field. One of the first approaches of this type was one in which the QCD vacuum was modeled as a dilute gas of instantons [6]. However, it was soon realized that large gluon fluctuations destroy the validity of these dilute gas models [7]. Since then, other attempts have been made such as modifying the QCD action functional [8] or proposing an instanton liquid [9]. To date, no one approach has been widely recognized as the correct way to derive phenomenological confinement models and parameters directly from QCD.

In this paper, we propose a twist to the background mean field approach. Instead of using mean field theory with a classical background gluon field, we propose using an operator background gluon field made up of quantized quark creation and destruction operators. In this fermionic background field approach, the background field is constructed so that its expectation value with respect to the vacuum vanishes. Due to this feature, the presence of a vector background field does not conflict with Lorentz invariance of the vacuum. With classical background fields, one cannot set the vacuum expectation value of the field equal to zero without setting the field as a whole equal to zero. With an operator field, however, it is clearly possible for a field with a vanishing vacuum expectation value to have a nontrivial structure that leads to other interesting physical consequences. The nontrivial structure of the field proposed here produces a color field energy term in the Hamiltonian that leads to an infinitely negative vacuum energy density in the quark sector of the theory. Since the quark mass matrix and coupling constant set the scale for the energy density, no phenomenological parameters are needed.

In mean field theory, the use of a mean field is justified by showing that the effect of gluon fluctuations around the mean field is small. Justification of the fermionic background field is done in a different way, by showing that gluon fluctuations completely decouple from quarks. This means that when considering ground states of various colorless quark configurations, one can assume that gluon fluctuations remain in the same minimum energy state, regardless of the quark configuration under consideration. Therefore, in energy differences between different ground-state configurations, the gluon fluctuation energy drops out, and the results are independent of the gluon fluctuation ground state. For example, when calculating hadron masses (the difference between hadron energy and vacuum energy), one can completely ignore gluon fluctuations and concentrate solely on a quark Hamiltonian involving the fermionic background gluon field.

A limitation of the present approach is that it can only be used for QCD and not for QED or the Weak interaction. We will show that the reason this approach cannot be used for QED is because it only applies to groups that have a traceless group generator. We will also show that the "diagonal" approach used in this paper only applies when fermion masses are independent of group indices. Since quark masses are independent of color, this restriction is met for QCD. However, since lepton and quark masses are not independent of the Weak SU(2) index, the condition is not met for the Weak interaction. It should be noted that a more general fermionic background field method can be made to work with fermions that have different masses for different group indices, but it requires the background field to be non-diagonal. This more complicated case will be addressed in a future work.

Since this paper does not address the Weak Interaction, the fields and 'vacuum' presented here do not exhibit chiral symmetry breaking. Although a true description of the vacuum must take into account this important physical concept, the symmetric model presented here provides a simpler presentation of the basic concepts of the fermionic background field method. Throughout the paper, we use the words "vacuum" and "hadron". These concepts should be understood to apply to a symmetric theory with no Weak Interaction. The follow-on work that addresses the Weak Interaction will show how the present description of the QCD vacuum is modified in the presence of chiral symmetry breaking.

Another limitation of the present approach is that it can only be used when considering matrix elements involving quantum states whose quarks are arranged in colorless combinations. We argue that the eigenstates of the vacuum and hadrons meet this restriction, so this method can be used to describe their properties. In its present form, the method is not well suited for describing perturbative QCD phenomena such as jets. However, since the method is based on the same Lagrangian as that used for perturbative QCD, it is not inconsistent with important concepts from that description such as a running coupling constant.

The purpose of the present paper is to present the fermionic background field method, to show how it leads to new nontrivial solutions to the quantum equations of motion, and to present the mechanism for decoupling of quarks and gluon fluctuations. Heuristic arguments are made as to how the background field can lead to stable hadron radii and masses. In addition to chiral symmetry breaking, true hadron mass calculations require quantization of quark fields inside a sphere rather than in plane waves. These calculations will be the topic of a coming paper, but in anticipation of them, we have attempted to prove as much of the method as possible in coordinate space before expanding fields in terms of plane waves at the end. The advantage of the coordinate space arguments is that they apply equally well to plane wave expansions and spherical expansions of the quark fields.

The outline of the paper is as follows: In the first section, we introduce assumptions of the fermionic background field method, identify the equations of motion and new form of the action involving the background field, and discuss the implications of a decoupling of quarks and gluon fluctuations. In the second section, we specify the form of the background field in coordinate space and prove most of the relations needed to show that the field solves the equations of motion and leads to quark-gluon decoupling in the Hamiltonian. At the end of this section, we show how the background field leads to negative energies that provide an explanation for confinement and a mechanism for establishing stable hadron radii and masses. In the third section, we introduce plane wave expansions of the fields, explicitly define the colorless quark states, and prove remaining identities that are delayed to this point. The section ends by showing that the remaining quark Hamiltonian is completely diagonalizable and renormalizable and has a form that validates a description of hadron interactions in terms of meson exchange.

**1. The fermionic background field method**

In the fermionic background field method, we seek quantized field configurations that form extrema of the QCD action within the context of colorless quark states. In other words, we seek solutions to

$$\langle x'|dS|x\rangle = 0. \tag{1.1}$$

Here $|x\rangle$ and $\langle x'|$ represent quantum states in which the pure quark part of the states is colorless. We will implicitly use canonical quantization along the lines of [10] rather than path integral quantization, since we will be working with solutions and field configurations that do not have a classical equivalent. In canonical quantization, one can think of the quantum states as being comprised of quark/anti-quark creation operators and gluon fluctuation creation operators acting on some vacuum state. Since the quark operators commute with the gluon fluctuation operators, these sectors of the Hilbert space can be treated independently. Therefore, a general state can be written

$$|x\rangle = \sum_{nm} c_{nm} |x_q^{(n)}\rangle |x_g^{(m)}\rangle, \tag{1.2}$$

where $|c_{nm}|^2 = 1$, $|x_q^{(n)}\rangle$ and $|x_g^{(m)}\rangle$ represent orthonormal bases of quark and gluon states, and we restrict the states $|x_q^{(n)}\rangle$ to colorless combinations of quark creation operators. Note that we are not restricting the gluon fluctuation states in any way.

In this paper, we assume that the QCD vacuum is colorless. In addition, we assume that the quarks in the vacuum separately form colorless combinations. As a result, we assume that the QCD vacuum eigenstate $|0\rangle$ is one of the states within $|x\rangle$, so the method presented here can be used to calculate the quark contribution to the QCD vacuum energy. Moreover, we assume that the pure quark part of a hadron eigenstate is colorless, so each hadron is one of the states in $|x\rangle$. Since hadron mass is the difference between hadron energy and vacuum energy, this method can be used to calculate hadron masses directly from QCD, without introducing phenomenological parameters. At the end of the paper, we argue that these assumptions are justified by the fact that colorless quark combinations lead to infinitely more negative energy that colored quark combinations.

The QCD action is given by:

$$S = \int d^4x \left( \overline{\Psi}(i\partial - M + gA)\Psi - \tfrac{1}{4} F^a_{mn} F^{mna} \right), \tag{1.3}$$

where $\Psi$ are the quark (and anti-quark) fields, $A_m^a$ are the gauge fields, and $M$ is the fermion mass matrix in color and flavor space. We use the usual notation where $A = A_m^a g^m T^a$ and $T^a$ are group matrices. For SU(3), we take $T^a = \lambda^a/2$ to be the Gell-Mann matrices.

In the context of colorless quark states, extrema of the QCD action satisfy the following equations

$$\langle x'|\int d^4x \overline{\Psi}(i\partial - M + g\overline{A})\Psi|x\rangle = 0 \tag{1.4}$$

$$\langle x'|\int d^4x \hat{A}_n^a \left[(\partial_m \delta^{ab} + gf^{abc}\overline{A}_m^c)\overline{F}^{mna} - g\overline{\Psi}T^a g^n \Psi\right]|x\rangle = 0, \tag{1.5}$$

where we denote gluon fields that solve (1.4) and (1.5) by $\overline{A}_m^a$ and fluctuations around these fields by $\delta A_m^a = \hat{A}_m^a$. We have also made the replacement

$$\delta \overline{\Psi}(x) = \overline{\Psi}(x). \tag{1.6}$$

The reason we can make this replacement is that quantized fields do not have a single value for a given $x$; the field can take any possible value that is consistent with the state the field acts on. Since a quantized field already has every possible variation in it, the field and its variations take the same form.

We look for solutions to the equations of motion (1.4) and (1.5) in which the background gluon field is diagonal in color space and made up of quantized quark operators. In other words, we seek solutions of the form:

$$\overline{A}_m^3 = \overline{A}_m^3(\Psi, \overline{\Psi}) \quad \text{and} \quad \overline{A}_m^8 = \overline{A}_m^8(\Psi, \overline{\Psi}), \tag{1.7}$$

where all "non-diagonal" background field components vanish. Since the quark fields are comprised of quark operators, the form of the solution proposed above is significantly different than those in more standard background field formalisms. Usually, background fields are assumed to be classical (non-operator) fields. In that case, for any given point of spacetime, the background field takes the same value no matter what quantum states act upon it. In the present case, because the background field is quantized and acts on the quark Hilbert space, the value of the field at a particular point of spacetime varies depending upon the states $|x_q^{(n)}\rangle$ acting upon it.

This quantum nature of the background field permits non-trivial solutions that preserve Lorentz invariance of the vacuum. To see this, we note that $\langle 0|\overline{A}^{ma}|0\rangle = 0$ is a sufficient condition to ensure that the presence of a vector background gluon field does not spoil Lorentz invariance of the vacuum. Although this condition is not necessary (instantons do not obey this condition), we will restrict the present analysis to this condition. For classical background fields, $\langle 0|\overline{A}^{ma}|0\rangle = 0$ implies $\overline{A}^{ma} = 0$, so there are no non-trivial solutions that obey this condition and preserve Lorentz invariance. However, since the background field in the present analysis is an operator, $\langle 0|\overline{A}^{ma}|0\rangle = 0$ does not imply $\overline{A}^{ma} = 0$, so it is possible to find non-trivial background field solutions that obey this condition. We will actually seek solutions that obey the stricter condition:

$$\langle x_q^{(n)}|\overline{A}^{ma}|x_q^{(m)}\rangle = 0. \tag{1.8}$$

Since the vacuum state $|0\rangle$ is one of the states in $|x\rangle$, the above condition implies $\langle 0|\overline{A}^{ma}|0\rangle = 0$, and Lorentz invariance of the vacuum is ensured.

We would like to consider the effect of fluctuations around the background field solution. To do this, we define the full gluon field by

$$A_m^a = \overline{A}_m^a + \hat{A}_m^a ,\tag{1.9}$$

where $\hat{A}^{ma}$ are quantized fluctuations that act on the gluon fluctuation states $\left|x_g^{(n)}\right\rangle$. Using (1.7) and (1.9), the QCD action can be rewritten as follows:

$$\begin{aligned}
S &= S_q + S_g + S_{qg}^{(1)} + S_{qg}^{(2)} + S_{qg}^{(3)} \\
S_q &= \int d^4x \left[ \overline{\Psi}(i\partial - M + g\overline{A})\Psi - \tfrac{1}{4}\overline{F}_{mn}^a \overline{F}^{mna} \right] \\
S_g &= \int d^4x \left( -\tfrac{1}{4}\hat{F}_{mn}^a \hat{F}^{mna} - \tfrac{1}{2}\partial_m \hat{A}^{ma} \partial_n \hat{A}^{na} \right) \\
S_{qg}^{(1)} &= -\tfrac{1}{2} g^2 f^{abc} f^{ade} \int d^4x \overline{A}_m^b \overline{A}^{ml} \hat{A}_n^c \hat{A}^{ne} \\
S_{qg}^{(2)} &= \tfrac{1}{4} \int d^4x \left\{ \left(d^{ab}\partial_m + igf^{abc}\overline{A}_m^c\right)\hat{A}^{mb}, \left(d^{ad}\partial_n - igf^{ade}\overline{A}_n^e\right)\hat{A}^{nd} \right\} \\
S_{qg}^{(3)} &= \int d^4x \left[ \overline{\Psi} g^m T^a \Psi g \hat{A}_m^a - \tfrac{1}{2}\overline{F}_{mn}^a \hat{F}^{mna} - g^2 f^{abc} \overline{A}_m^b \hat{A}_n^c \hat{F}^{mna} \right] .
\end{aligned}\tag{1.10}$$

We have split the total action into a quark-only action $S_q$, a gluon fluctuation-only action $S_g$, and three terms involving quark-gluon interactions. In order to get $S_{qg}^{(2)}$, we have used the Jacobi identity and have included a pure-gluon derivative term that is also subtracted off of $S_g$. Although this action looks very complicated, we will show that the Hamiltonian created from it simplifies considerably.

In fact, we will show that the background field and the gauge for fluctuations can be chosen so that

$$\left\langle x' | H_{qg} | x \right\rangle = \left\langle x' | \left( H_{qg}^{(1)} + H_{qg}^{(2)} + H_{qg}^{(3)} \right) | x \right\rangle = 0 .\tag{1.11}$$

In other words, in the context of colorless quark matrix elements, the quark-gluon interaction terms vanish, and the quark and gluon parts of the Hamiltonian decouple. Given this decoupling, a general matrix element of the Hamiltonian takes the form:

$$\left\langle x' | H | x \right\rangle = \sum_{nmp} c'_{np} c_{mp} \left\langle x_q^{(n)} | H_q | x_q^{(m)} \right\rangle + \sum_{nmp} c'_{pn} c_{pm} \left\langle x_g^{(n)} | H_g | x_g^{(m)} \right\rangle .\tag{1.12}$$

For any given quark configuration, the energy is minimized if the gluon fluctuations $\left|x_g^{(m)}\right\rangle$ are in the lowest energy eigenstate of $H_g$. Since quarks and gluon fluctuations are decoupled, the form of this gluon fluctuation eigenstate is independent of the particular quark states under consideration. When comparing lowest energy states of different colorless quark combinations, we can assume the gluon fluctuations are in this ground state, so we have

$$\left\langle x' | H_{eff} | x \right\rangle = \left\langle x'_q | H_q | x_q \right\rangle + \left\langle x_g^{(0)} | H_g | x_g^{(0)} \right\rangle \tag{1.13}$$

where

$$\left|x_q\right\rangle = \sum_n c_{n0} \left|x_q^{(n)}\right\rangle \tag{1.14}$$

and $\left|x_g^{(0)}\right\rangle$ is the ground state of the pure gluon Hamiltonian.

Equation (1.13) has an interesting implication: Suppose $|h\rangle = |h_q\rangle |x_g^{(0)}\rangle$ is a hadron eigenstate of the full Hamiltonian $H$. If the quark state $|h_q\rangle$ within $|h\rangle$ is colorless, then the

energy of the hadron is given by equation (1.13). This means that the mass of the hadron is given by

$$m_h = \langle h_q | H_q | h_q \rangle - \langle 0 | H_q | 0 \rangle \tag{1.15}$$

and is independent of the gluon fluctuation energy. In other words, one can completely ignore gluon fluctuations when calculating hadron masses.

## 2. **The background gluon field**

In this section, we explicitly define the background gluon field in terms of quark fields then show how it solves the equations of motion and leads to quark-gluon decoupling in the Hamiltonian. Some of the needed relations require momentum expansions of the quark fields, but we delay these to the next section, taking the coordinate analysis as far as possible in this section. A subtlety of the method arises due to the fact that the background field is dependent on a 4-velocity $u$ with regard to some unspecified reference frame. To show that this dependence is not a problem, we verify that none of the equations of motion depend on $u$, so $u$ can be taken to be any convenient frame. However, we also show that quarks and gluon fluctuations only decouple in the Hamiltonian if we take $u$ to be the reference frame chosen for calculation of the Hamiltonian.

We now propose the following form for the background gluon field:

$$g\overline{A}_m^a(x) = i\int d^4y\, \mathbf{d}(u \cdot (x-y)) \partial_m \overline{c}(y) u \cdot \mathbf{g} M \overline{T}^a \Psi(y) \tag{2.1}$$

where $u$ is the 4-velocity of some as-yet-undetermined reference frame, $\overline{c}$ is a "generating" quark field defined through

$$\overline{\Psi} = -i\overline{c}\overleftarrow{\partial} = -i\partial_m \overline{c}\mathbf{g}^m, \tag{2.2}$$

and we have defined SU(3) diagonal group matrices

$$\overline{T}^a = \tfrac{3}{2}\left(\mathbf{d}^{a3}\mathbf{1}^3 + \mathbf{d}^{a8}\mathbf{1}^8\right) \tag{2.3}$$

that satisfy $\overline{T}^a T^a = 1$. Two very interesting attributes of the background field are that it is nonlocal and that it depends some reference frame $u$. We will show below that the nonlocal integral form of the background gluon field allows it to satisfy an important commutation relation that is needed to solve the equations of motion. One could heuristically think of this nonlocal behavior as being the result of summing over an infinite number of perturbation theory diagrams involving multiple derivatives. The dependence on a reference frame $u$ appears to be a problem, but as stated above, we will show below that none of the equations of motion depend on $u$, so this reference frame does not violate the relativity principle.

It is important that the background gluon field transforms as a Lorentz 4-vector. One way to see that this is the case is from the definition of $\overline{c}$ in (2.2) and the fact that $\partial$ is a Lorentz invariant. From these, it follows that $\overline{c}$ has the same Lorentz transformation properties as $\overline{\Psi}$, so $\partial_m \overline{c} u \cdot \mathbf{g}\Psi$ is a 4-vector. Another way to see this is by rewriting the background field as follows:

$$g\overline{A}_m^a(x) = i\int d^4y\, \mathbf{q}(u \cdot (x-y))\partial_n\left[\partial_m \overline{c}(y)\mathbf{g}^n M\overline{T}^a \Psi(y)\right]$$

$$= -\int d^4y\, \mathbf{q}(u \cdot (x-y))\left[\partial_m \overline{\Psi}(y) M \overline{T}^a \Psi(y) - i\partial_m \overline{c}(y) M \overline{T}^a \partial\Psi(y)\right], \tag{2.4}$$

where we have assumed that quark fields vanish at spacelike infinity (relative to $u$) and have neglected a term at time-like negative infinity that is independent of $x$, since it shares transformation properties with the terms shown above. Again, since $\overline{c}\partial$ transforms like $\Psi$, both terms in the second line of (2.4) are Lorentz 4-vectors.

To see how the background field solves the quantum Dirac equation of motion (1.4), we start by noting that a Lorentz-invariant generalization of the usual equal time fermion field anticommutation relations is:

$$\{\overline{\Psi}(x), u \cdot \gamma \Psi(y)\} \delta(u \cdot (x-y)) = \delta^4(x-y), \qquad (2.5)$$

for any 4-velocity $u$. Using this relation along with $\{\overline{\Psi}(x), \overline{c}(y)\} = 0$ and the definition of $\overline{c}$ in (2.2), we have:

$$\int d^4x [\overline{\Psi}, g\overline{A}_m^a] T^a \gamma^m \Psi = \int d^4x \overline{\Psi} M \Psi. \qquad (2.6)$$

Using this commutation relation, the quantum Dirac equation simplifies to:

$$\langle x'| \int d^4x \overline{\Psi}(i\partial - M + g\overline{A}) \Psi |x\rangle = \langle x'| \int d^4x (\overline{\Psi} i\partial \Psi + g\overline{A}_m^a J^{ma}) |x\rangle = 0, \qquad (2.7)$$

where $J^{ma} = \overline{\Psi} \gamma^m T^a \Psi$. The Dirac equation can then be solved for all values of g, if

$$i\partial \Psi = 0 \qquad (2.8)$$

$$\langle x'| \int d^4x g\overline{A}_m^a J^{ma} |x\rangle = 0. \qquad (2.9)$$

Since equation (2.5) is independent of the reference frame $u$, the above solution to the quantum Dirac equation is independent of $u$ as long as (2.9) is independent of $u$. The latter will be shown in the next section.

Let us make a few observations about this solution. First, since equation (2.6) involves a commutator, the solution presented here is a quantum solution that does not have a classical analog. This commutator term exactly cancels the quark mass term in the Lagrangian, thereby allowing the quark fields to obey the massless equation of motion (2.8). In the next section, we will show that equation (2.9) is only satisfied for quark fields that obey massless equations of motion, so cancellation of the quark mass term by the commutator is important. This cancellation sets the scale of the background field by the quark mass matrix in the Lagrangian, so no phenomenological parameter is needed to set that scale. It is also now apparent that the nonlocal space-like integral of the background field is needed in order to offset the space-like delta function that arises in the quark field anti-commutator (2.5).

We have shown how the background field satisfies the quantum Dirac equation (1.4). Before showing how the proposed background field satisfies the other equation of motion (1.5), we first examine the condition (1.8) that ensures the Lorentz invariance of the vacuum. In the next section, we will define our colorless quark states so that they involve sums over color indices and treat no color different from any other. This being the case, the matrix element $\langle x_q^{(n)} | \overline{A}^{ma} | x_q^{(m)} \rangle$ results in a color trace of the matrix $M\overline{T}^a$. If the quark mass matrix is independent of color, then the trace is over the traceless group matrix $\overline{T}^a$, and the matrix element vanishes. From these arguments, it can be seen that the present method would not work for QED (since the group generator is not traceless) or for the Weak Interaction (since quark and lepton masses are not independent of the SU(2) index). The present analysis only works for QCD due to the facts that the group generators are traceless and quark masses are independent of color.

Since the background gluon field is diagonal, $\overline{F}_{mn}^a = \partial_m \overline{A}_n^a - \partial_n \overline{A}_m^a$, and the above arguments can also be used to see that

$$\left\langle \mathbf{x}_q^{(n)} \middle| \overline{F}_{mn}^a \middle| \mathbf{x}_q^{(m)} \right\rangle = 0$$
$$\left\langle \mathbf{x}_q^{(n)} \middle| J^{ma} \middle| \mathbf{x}_q^{(m)} \right\rangle = 0. \tag{2.10}$$

As a result of these equations and the fact that $f^{abc} \overline{A}_n^b \overline{F}^{mnc} = 0$ for a diagonal background field, the "gluon" equation of motion (1.5) is satisfied. Because the needed relations are only dependent upon the color structure of the background field and color traces, they are valid for any reference frame $u$. Since the background field of (2.1) exactly solves both equations of motion (1.4) and (1.5) for any reference frame, we are free to pick any frame that is convenient for calculations.

Calculation of the Hamiltonian in a general reference frame is complicated due to the fact that the quark fields within one background field factor may have different time coordinates from those within another background field factor or those not inside a background field. However, in the frame defined by $u = (1,0,0,0)$, all fields in the Lagrangian have the same time coordinate whether or not they are inside a background field, so quantum field theory can be reduced to quantum mechanics through the usual method of treating the time-dependent field at each point of space as a separate canonical variable. Thus, the Hamiltonian simplifies considerably if one takes the frame $u$ to be the reference frame used for calculation of the Hamiltonian. Since the equations of motion are independent of $u$, one is free to make this choice.

Before calculating the Hamiltonian, we must decide on which variables to use as the canonical variables. The usual choice for fermions is to take $\Psi$ and $\overline{\Psi}$ as the canonical variables. However, in this case since the background field depends explicitly on $\overline{c}$ rather than $\overline{\Psi}$, the expressions are greatly simplified if we use $\overline{c}$ as our second canonical variable instead of $\overline{\Psi}$. This choice does not present any new complications, since the action is still only dependent on first derivatives of $\overline{c}$ and no second derivatives (we show this shortly). As a result, the usual canonical formalism can be used throughout.

We begin by calculating $H_{qg}^{(1)}$, which is the Hamiltonian derived from $S_{qg}^{(1)}$. Consider the following in the "Hamiltonian frame" $u = (1,0,0,0)$:

$$\frac{d g \overline{A}_m^a(y)}{d(\partial_0 \overline{c})} = i d_{m0} d(x_0 - y_0) g^0 M \overline{T}^a \Psi(y) \tag{2.11}$$

From this expression, it follows that $H_{qg}^{(1)}$ is given by the simple expression:

$$H_{qg}^{(1)} = -\tfrac{1}{2} g^2 f^{abc} f^{ade} \int d^3 x \left( \overline{A}_0^b \overline{A}_0^d + \overline{A}_i^b \overline{A}_i^d \right) \hat{A}_n^c \hat{A}^{ne}. \tag{2.12}$$

In the next section, by using momentum expansions, we will show explicitly that

$$\left\langle \mathbf{x}_q^{(n)} \middle| H_{qg}^{(1)} \middle| \mathbf{x}_q^{(m)} \right\rangle = 0. \tag{2.13}$$

For now, we note that the reason that (2.13) is satisfied is because the temporal background field is Hermitian and the spatial background field is anti-Hermitian, so the sum of their squares cancels for massless quark fields. The fact that the spatial background field is anti-Hermitian can be most easily seen by taking the conjugate of the first term of (2.4). A spatial integration by parts returns the term to its original form multiplied by minus one, so it is anti-Hermitian. Since the second term of (2.4) vanishes using the quark equations of motion, the fact that the spatial components of the first term of (2.4) are anti-Hermitian means that the spatial components of the entire background field are anti-Hermitian. This anti-Hermitian nature of the spatial background gluon field is also important both for gauge fixing and for establishing the negative energy of the vacuum.

Next we consider $H_{qg}^{(3)}$. Every term of $S_{qg}^{(3)}$ involves a single factor of $\bar{c}$ (or its derivatives) and a single factor of $\Psi$ (or its derivatives) that surround a factor of $T^a$. As stated before, these kinds of terms effectively take a trace of $T^a$, so they vanish. Moreover, the canonical momentum terms derived from this part of the action also have the same form, so they vanish as well. In other words $\langle \mathbf{x}_q^{(n)} | H_{qg}^{(3)} | \mathbf{x}_q^{(m)} \rangle = 0$, as a result of the fact that $T^a$ are traceless and quark masses are independent of color.

Finally we consider $H_{qg}^{(2)}$. We note that it is possible to pick a gauge condition for the gluon fluctuations that causes this part of the interaction Hamiltonian to vanish. The gauge we pick is a state-dependent combination of the temporal and background Coulomb gauge in the reference frame $u$:

$$\hat{A}_0^a = 0$$
$$\left( \mathbf{d}^{ab}\partial_i + gf^{abc}\bar{a}_i^c \right)\hat{A}^{ib} = 0, \tag{2.14}$$

Here $\bar{a}_i^a$ are Hermitian classical fields whose spatial components (in the frame $u$) are derived from the background gluon field via the relation:

$$\langle \mathbf{x}_q^{(n)} | \bar{A}_i^a \bar{A}_j^b | \mathbf{x}_q^{(m)} \rangle = \left( i\bar{a}_i^a \right)\left( i\bar{a}_j^b \right). \tag{2.15}$$

Note that the $\bar{a}_i^a$ are only Hermitian as a result of the fact that the $\bar{A}_i^a$ are anti-Hermitian. Since the $\bar{a}_i^a$ are classical Hermitian fields, the usual background gauge formalism [11] can be applied.

By looking at the form of the action in (1.10), it can be seen that in the Hamiltonian frame $u = (1,0,0,0)$, the temporal/background Coulomb gauge causes $\langle \mathbf{x}_q^{(n)} | S_{qg}^{(2)} | \mathbf{x}_q^{(m)} \rangle = 0$. As an aside, by generalizing the gauge condition through expressions such as $u^m \hat{A}_m^a = 0$, one can cause this part of the action to vanish for any frame $u$. This result could have been achieved more simply by using a background Lorentz gauge without giving the temporal component special treatment. The reason for using a temporal/background Coulomb gauge is because the canonical momentum term in $H_{qg}^{(2)}$ derived from $S_{qg}^{(2)}$ involves terms that do treat the temporal component of the gluon field differently. Since each of these terms is multiplied by one or more factors of $\hat{A}_0^a$, the temporal part of the gauge condition ensures that they vanish independently. In other words, in the gauge identified by (2.14) and (2.15), $\langle \mathbf{x}_q^{(n)} | H_{qg}^{(2)} | \mathbf{x}_q^{(m)} \rangle = 0$.

Putting together the results, we have shown that it is possible to pick a gauge for gluon fluctuations and a reference frame for the background gluon field such that the entire interaction Hamiltonian vanishes in any colorless quark matrix element. Hence, quarks and gluon fluctuations completely decouple.

As stated in the last section, we will assume that gluon fluctuations have found their lowest energy state and focus solely on the quark sector. Since (2.6) is valid simply as a result of field commutation relations and does not require the equations of motion (2.8), the quark mass term cancellation can be done before calculation of the Hamiltonian. In other words, we start with the quark action:

$$S_q = \int d^4x \left[ \overline{c} \overline{\partial} \partial \Psi + g \overline{A}_m^a J^{ma} - \tfrac{1}{4} \overline{F}_{mn}^a \overline{F}^{mna} \right], \tag{2.16}$$

where we have used (2.2) to rewrite $\overline{\Psi} \partial \Psi$. It is easily verified that the first two terms of (2.16) lead to $-\int d^3x \left( \overline{\Psi} i g^i \partial_i \Psi + g \overline{A}_i^a J^{ia} \right)$ in the Hamiltonian. Note that the contribution to the momentum canonical to $\overline{c}$ generated by the term $\overline{c} \overline{\partial} \partial \Psi$ vanishes due to the equations of motion (2.8). To complete the quark Hamiltonian, it is necessary to calculate the contribution from the field energy term.

Fortunately, in the Hamiltonian frame, the field energy term simplifies considerably. We have:

$$-\tfrac{1}{4} \overline{F}_{mn}^a \overline{F}^{mna} = \tfrac{1}{2} (\partial_0 A_i^a)^2 = -\frac{1}{2g^2} \left[ \int d^3y \left( \partial_i c^+ M \overline{T}^a \partial_0 \Psi - \partial_0 c^+ M \overline{T}^a \partial_i \Psi \right) \right]^2, \tag{2.17}$$

where we have used the fact that spatial surface terms vanish. The above expression is useful since it shows that the background field strength only depends on first derivatives of the quark fields. Since the kinetic energy term $\overline{\Psi} \partial \Psi$ also only depends on first derivatives, the action as a whole only depends on quark fields and their first derivatives. This was part of the justification for using $\overline{c}$ as one of the canonical variables.

Due to the presence of the time derivatives in (2.17), a canonical momentum term contributes to the background field energy part of the quark Hamiltonian. This canonical momentum term is twice the magnitude of the field energy term in the Lagrangian, and results in the following quark Hamiltonian:

$$\begin{aligned} H_q &= -\int d^3x \left( \overline{\Psi} i g^i \partial_i \Psi + g \overline{A}_i^a J^{ia} \right) + \tfrac{1}{2} \left( \int d^3x \right) \left( \partial_0 \overline{A}_i^a(t) \right)^2 \\ &= \int d^3x \, \overline{\Psi}^+ i \partial_0 \Psi + \frac{1}{2g^2} \left( \int d^3x \right) \left( \int d^3y \, \overline{\Psi} M \overline{T}^a \partial_i \Psi \right)^2. \end{aligned} \tag{2.18}$$

To get the second line, we have used (2.4), the quark equation of motion, and the fact that the current term vanishes in colorless quark matrix elements (as we show in the next section). The Hamiltonian of (2.18) is interesting in that in addition to the usual quark kinetic energy term, there is a background electric field energy term. As the square of an anti-Hermitian operator, it can be seen that the field energy term is negative. We now discuss some of the interesting implications of this negative field energy.

It is well known that the usual perturbative formulations of quantum field theories result in a vacuum with a large positive energy density [12,13]. This energy density is the result of divergent gluon loops and is valid for QCD as long as quarks couple to gluons and gluons couple to each other. In this paper, we have shown a method whereby quarks and gluons decouple for the special case of colorless quark states. Although colorless quark states may still lead to positive vacuum energies in the pure gluon sector of the theory, (2.18) shows that they lead to negative energies in the quark sector. As a result, the overall energy of colorless quark states in the present fermionic background field formalism is infinitely more negative than any quark state in the usual perturbative formalism. In the absence of another exact, negative-energy solution to the equations of motion that is applicable for colored quark states, the perturbative formalism must be used for these states. This suggests that colored quark states have infinitely more energy than colorless quark states, since the negative-energy fermionic background field formalism can only be used on the latter. This heuristic explanation for confinement implies that both the vacuum and any other physically observable quark states must involve only colorless combinations of quarks.

Even within these colorless combinations, there are restrictions. Since the electric field energy is only dependent on time, the integral $\int d^3x$ just becomes a spatial volume factor. This volume factor applies to the space that is relevant to the quantization of the quark fields that is defined via (2.5). This quantization in turn depends upon fields either vanishing or being periodic on the spatial surface of the volume under consideration. For example, if one is considering the usual plane wave expansions of quark fields, the applicable volume factor is infinite. This means that it would take infinite energy to create a colorless combination of plane-wave quark fields on top of the vacuum. However, if one quantizes the fields inside a sphere with fields that vanish on the surface, then the applicable volume factor would be the volume of the sphere. Since in these cases the volume is finite, the energy needed to create such states would also be finite.

It is now apparent how the field energy term of (2.18) leads to stable hadron radii and masses. Inside a sphere of radius $R$, one can make the replacement $i\partial_0 \to \boldsymbol{p}/R$ (for the ground state), so the mass of the hadron $h$ must be proportional to

$$\langle h|H|h\rangle - \langle 0|H|0\rangle \approx \frac{\boldsymbol{p} N_q}{R} + \frac{2\boldsymbol{p} R^3 M^2}{3g^2}\frac{f(h)}{R^2}, \qquad (2.19)$$

where we have extracted an $R^2$ out of the matrix element of the field energy to leave a dimensionless function $f(h)$. We can see that minimizing this mass will lead to a stable hadron radius proportional to $g/M$, where $M$ is some combination of the various quark masses in the hadron. It should be noted that this method of calculating hadron masses does not depend upon phenomenological parameters such as constituent quark masses, but rather on the current quark masses in the QCD Lagrangian. Spherical quantization and hadron mass calculations will be addressed more fully in a coming paper, but in the meantime, we would like to mention that the Hamiltonian of (2.18) reproduces key elements of the hadron mass spectrum, such as spin-dependent masses.

### 3. Quark Field Expansions

In this section, we begin by defining momentum expansions of the quark fields that obey the massless equations of motion. Using these to define a color charge operator, we explicitly define the colorless quark states so that they cause the color charge operator to vanish. We show that this feature validates (2.9) and the fact that the proposed background gluon field forms an exact solution to the quantum Dirac equation, independent of the reference frame $u$. Next we use the explicit colorless quark states to prove that $H_{qg}^{(1)}$ of (2.12) does indeed vanish in colorless quark matrix elements. This is the last piece needed to show that quarks and gluon fluctuations decouple in this theory. Finally, we show how the quark Hamiltonian of (2.18) is consistent with an interpretation of hadron interactions in terms of meson exchange.

We define the following plane-wave expansions of the quark fields:

$$\Psi = \sum_{\boldsymbol{a}sf} C_{\boldsymbol{a}} F_f \int \frac{d^3 p}{p^0 \sqrt{2(2\boldsymbol{p})^3}} \begin{pmatrix} p^0 S_s \\ \vec{p}\cdot\vec{\boldsymbol{s}}\ S_s \end{pmatrix} \left(b_{\boldsymbol{a}sfp} e^{-ip\cdot x} + d^+_{\boldsymbol{a}sfp} e^{ip\cdot x}\right), \qquad (3.1)$$

$\bar{s}$ to represent the spin opposite to *s*. We have taken $p^0 = |\vec{p}|$, assumed the Dirac representation of the gamma matrices, and have defined $C_a$, $F_f$, and $S_s$ as unit vectors in color, flavor, and spin space.

It is easily verified that the fields defined in (3.1) satisfy the equation $i\partial\Psi = 0$. By imposing the usual anticommutation relations on the amplitudes

$$\{b_{asfp}, b^+_{a's'f'p'}\} = \{d_{asfp}, d^+_{a's'f'p'}\} = \delta_{aa'}\delta_{ss'}\delta_{ff'}\delta^3(\vec{p}-\vec{p}'), \qquad (3.2)$$

one also recovers the Lorentz-invariant field anti-commutator of (2.5). Next we define a state that satisfies

$$b_{asfp}|x_0\rangle = d_{asfp}|x_0\rangle = 0. \qquad (3.3)$$

This allows us to interpret the amplitudes as creation and destruction operators relative to that state. Note that we are not saying that $|x_0\rangle$ is the true vacuum, but it will be needed to define the true vacuum and hadron states.

In the last section, we stated that the current term in the quantum Dirac equation vanishes. Now we show how this occurs. Since the quantity $\int d^4 x g A^a_m J^{ma}$ is a Lorentz invariant, it takes the same value in any reference frame. Calculation is easiest in the Hamiltonian frame where the background field is independent of spatial coordinates, so the integral over $d^3x$ only acts on the current $J^{ma}$. Consequently, we define the following color charge vector:

$$Q^{ma} = \int d^3x J^{ma} = Q^{ma(1)} + Q^{ma(2)}$$

$$Q^{ma(1)} = \int d^3x J^{ma} = \sum_{asf} T^a_a \int \frac{d^3p \, p^m}{p_0}\left(b^+_{asfp}b_{asfp} - d^+_{asfp}d_{asfp}\right)$$

$$Q^{ma(2)} = \int d^3p Q^{ma(2)}_p = i\delta^{mi} \sum_{ass'f} T^a_a \int \frac{d^3p}{p^0} S^T_{s'}\epsilon^{ijk} p^j S^k S_s \left(d_{as'f\bar{p}}b_{asfp}e^{-2ip^0t} - b^+_{asfp}d^+_{as'f\bar{p}}e^{2ip^0t}\right), \qquad (3.4)$$

where we use the notation $\bar{p}^m = p_m = (p^0, -p^i)$. Since the current in the Dirac equation is multiplied by a diagonal background gluon field, we restrict our analysis to charges with diagonal group matrices and use the notation $T_a = C^T_a T^a C_a$.

The following condition,

$$\langle x_q^{(n)} | g A^a_m Q^{ma} | x_q^{(m)} \rangle = 0, \qquad (3.5)$$

is sufficient to satisfy equation (2.9) and hence solve the quantum Dirac equation. From (3.4), it is easily seen that the time-dependent part of the charge operator satisfies the relation:

$$p_i Q^{ia(2)}_p = 0. \qquad (3.6)$$

We will show shortly that this is exactly what is needed to make the part of (3.5) involving the time-dependent charge vanish. For the present, we will focus on the time-independent charge term and look for colorless quark states that satisfy

$$Q^{ma(1)}|x_q^{(n)}\rangle = 0. \qquad (3.7)$$

For the spatial components of the above equation to apply, the quark states must be separately colorless for each momentum index.

Keeping this in mind, we define the following colorless state creation operators:

$$B^+_{ss's''ff'f''p} \equiv \frac{1}{\sqrt{6}} \sum_{abg} e_{abg} b^+_{asfp} b^+_{bs'f'p} b^+_{gs''f''p}$$

$$\overline{B}^+_{ss's''ff'f''p} \equiv \frac{1}{\sqrt{6}} \sum_{abg} e_{abg} d^+_{asfp} d^+_{bs'f'p} d^+_{gs''f''p}$$

$$M^+_{ss'ff'p} \equiv \frac{1}{\sqrt{3}} \sum_{a} b^+_{asfp} d^+_{as'f'p} , \qquad (3.8)$$

where we will call these operators baryon, anti-baryon, and meson operators, or collectively "hadron" operators. Notice that in all of these operators, every quark or anti-quark has the same momentum index, although they can have different spin and flavor indices.

These operators satisfy

$$\left[Q^{ma(1)}, B^+_{ss's''ff'f''p}\right] = \left[Q^{ma(1)}, \overline{B}^+_{ss's''ff'f''p}\right] = \left[Q^{ma(1)}, M^+_{ss'ff'p}\right] = 0. \qquad (3.9)$$

The baryon and anti-baryon commutators accomplish this by taking a trace of the color matrix, while the meson commutator vanishes due to the fact that $Q^{ma(1)}$ also measures baryon number for each color/momentum combination. We now explicitly define the colorless quark states $\left|x_q^{(n)}\right\rangle$ to be states made from any combination of the hadron creation operators of (3.8) acting on the state $\left|x_0\right\rangle$. Due to the commutation relations (3.9), these states satisfy condition (3.7), causing the time-independent charge term to vanish.

As for the time-dependent term, since every state in (3.5) is colorless for every momentum index, the operator $gA_i^a Q^{ia(2)}$ will only contribute to the matrix element for terms that are colorless for each momentum index. Because the quark operators within $Q^{ia(2)}$ have opposite momentum indices, it is only possible for $gA_i^a Q^{ia(2)}$ to be colorless for a given momentum when both $gA_i^a$ and $Q^{ia(2)}$ share the same momentum index. However, since $gA_i^a \propto p_i$ in momentum space (this is shown explicitly below), we can see from (3.6) that $gA_i^a Q^{ia(2)}$ vanishes when both factors share the same momentum index. In short, the time-dependent charge part of (3.5) vanishes due to the fact that our quark states are colorless for each momentum index.

We have shown that the current term in the quantum Dirac equation and the Hamiltonian vanishes when calculated in the Hamiltonian frame. However, since the quantity $\int d^4x g A_m^a J^{ma}$ is a Lorentz invariant, the fact that the current term vanishes in the Hamiltonian frame means that it vanishes in any frame. In other words, we have finished showing that the equations of motion (1.4) and (1.5) are satisfied by the background field, independent of the reference frame $u$.

We mentioned in the last section that the current term would not vanish if the quarks obeyed a massive field expansion. It is straightforward to verify that in this case, the spatial charge $Q^{ia}$ would have a term proportional to $S_s^T s^i S_s$. This term does not cause (3.5) to vanish, so the quantum Dirac equation is not solved. In other words, the quark fields must obey massless equations of motion in order for the current term to vanish. As stated earlier, this restriction is important since it sets the scale of the background field by that of the quark mass matrix in the Lagrangian through the relation (2.6).

Next we will finish showing that quarks and gluon fluctuations decouple in the Hamiltonian frame. To do this we first need to write down the form of the generating quark field:

$$c = \sum_{asf} C_a F_f \int \frac{d^3p}{2p_0^2 \sqrt{2(2\mathbf{p})^3}} \begin{pmatrix} p^0 S_s \\ -\vec{p}\cdot\vec{S}\ S_s \end{pmatrix} \left( b_{asfp} e^{-ip\cdot x} - d^+_{asfp} e^{ip\cdot x} \right) \quad (3.10)$$

This field can be seen to satisfy $\Psi = i\partial c$, which is just the conjugate of (2.2).

Using (3.1) and (3.10), the background gluon field in the Hamiltonian frame can be seen to have the expansion:

$$g\overline{A}^a_m = -\sum_{asf} M_f T^a_a \int \frac{d^3p}{2p_0} \left( \overline{p}_m d_{as\overline{f}p} b_{asfp} e^{-2ip^0 t} + p_m b^+_{asfp} d^+_{as\overline{f}p} e^{2ip^0 t} \right), \quad (3.11)$$

where we use the notation $M_f = F_f^T M F_f$ (assuming that the flavor vectors diagonalize the quark mass matrix). From (3.11), it is apparent that the temporal components of the background field are Hermitian, but the spatial components are anti-Hermitian. This difference is what allows cancellation in (2.13) and leads to decoupling of quarks and gluon fluctuations.

To see explicitly how this cancellation comes about, let us first consider a product of spatial fields:

$$g^2 \langle \mathbf{x}_q^{(n)} | \overline{A}_i^a \overline{A}_i^a | \mathbf{x}_q^{(m)} \rangle = -\tfrac{1}{4} \sum_{ass'ff'} M_f M_{f'} \overline{T}^a_a T^a_a \int (d^3p)(d^3p) \langle \mathbf{x}_q^{(n)} | \left[ \{ d_{as'\overline{f'}\overline{p}} b_{as'f'p}, b^+_{asfp} d^+_{as\overline{f}\overline{p}} \} + \right.$$

$$\left. + d_{as'\overline{f'}\overline{p}} b_{as'f'p} d_{as\overline{f}\overline{p}} b_{asfp} e^{-4ip^0 t} + b^+_{as'f'p} d^+_{as'\overline{f'}\overline{p}} b^+_{asfp} d^+_{as\overline{f}\overline{p}} e^{4ip^0 t} \right] | \mathbf{x}_q^{(m)} \rangle. \quad (3.12)$$

Because the states $\langle \mathbf{x}_q^{(n)} |$ and $| \mathbf{x}_q^{(m)} \rangle$ are separately colorless for each momentum state, the only contribution to the operator $\overline{A}_i^a \overline{A}_i^a$ comes from terms that are separately colorless for each momentum state. This fixes the momentum of the second factor of $\overline{A}_i^a$ to be the same as that of the first factor of $\overline{A}_i^a$ for the time-independent terms of (3.12) and opposite to that of the first factor of $\overline{A}_i^a$ for the time-dependent terms of (3.12). In continuum space, it is awkward to fix the momenta in this way and the result is the strange looking factor $\int (d^3p)(d^3p)$. This factor is best understood by moving from continuous space to discretized space inside some large finite volume. In this case, integrals over momentum are replaced by sums, and quark creation and destruction operators are dimensionless rather than having dimension $-\tfrac{3}{2}$. We will use the notation $\int (d^3p)(d^3p)$ to represent the quantity below that is well defined in discretized space:

$$\int (d^3p)(d^3p) = \int d^3p \int_{q=p} d^3q \to \sum_{pq} d_{pq}. \quad (3.13)$$

Using the same reasoning as above, it is easily verified that a product of temporal components of the background field gives the same result as (3.12), except with the opposite sign. This leads to the relation:

$$g^2 f^{abc} f^{ade} \langle \mathbf{x}_q^{(n)} | \left( \overline{A}_0^b \overline{A}_0^d + \overline{A}_i^b \overline{A}_i^d \right) | \mathbf{x}_q^{(m)} \rangle \int d^3x \hat{A}_n^c \hat{A}^{ne} =$$

$$= g^2 \langle \mathbf{x}_q^{(n)} | \left( \overline{A}_0^a \overline{A}_0^a + \overline{A}_i^a \overline{A}_i^a \right) | \mathbf{x}_q^{(m)} \rangle \int d^3x \left( \hat{A}_n^b \hat{A}^{nb} - \hat{A}_n^3 \hat{A}^{n3} - \hat{A}_n^8 \hat{A}^{n8} \right) = 0, \quad (3.14)$$

where we have used the fact that $\left\langle \mathbf{x}_q^{(n)} \middle| \overline{A}_m^3 \overline{A}_n^8 \middle| \mathbf{x}_q^{(m)} \right\rangle = 0$, since $\overline{T}^3 \overline{T}^8$ is a traceless matrix. Equation (3.14) means that in the Hamiltonian frame, quarks and gluon fluctuations decouple. As a result, we have validated that the Hamiltonian in the context of colorless quark states takes the simple form of (2.18).

Now we would like to use the Hamiltonian of (2.18) to calculate the energy of the vacuum and of other colorless quark configurations. To do this, we start by writing down the momentum expansion of the field strength:

$$-g\partial_0 \overline{A}_i^a = \int d^3 y \overline{\Psi} M \overline{T}^a \partial_i \Psi = i \sum_{asf} M_f T_a^a \int d^3 p\, p_i \left( d_{\mathbf{a}s\bar{f}p} b_{\mathbf{a}sfp} e^{-2ip^0 t} + b_{\mathbf{a}sfp}^+ d_{\mathbf{a}s\bar{f}p}^+ e^{2ip^0 t} \right). \quad (3.15)$$

Just as in the preceding analysis, in a product of field strength factors, the only contribution comes from terms in which both factors have the same or opposite momenta. If these terms are put into normal order, the resulting anti-commutators lead to additional diagonal terms.

Putting everything together, we find:

$$\left\langle \mathbf{x}_q^{(n)} \middle| H_q \middle| \mathbf{x}_q^{(m)} \right\rangle = \left\langle \mathbf{x}_q^{(n)} \middle| (H_0 + H_I) \middle| \mathbf{x}_q^{(m)} \right\rangle$$

$$H_0 = \sum_{asf} \int d^3 p\, p^0 \left( 1 + \frac{3VM_f^2 p^0}{2g^2} \right) \left( b_{\mathbf{a}sfp}^+ b_{\mathbf{a}sfp} + d_{\mathbf{a}sfp}^+ d_{\mathbf{a}sfp} - 1 \right)$$

$$H_I = -\frac{3V}{2g^2} \sum_{ass'ff'} M_f M_{f'} \int (d^3 p)(d^3 p) p_0^2 \times$$

$$\times \left( (d_{\mathbf{a}s'\bar{f}\bar{p}} b_{\mathbf{a}sf\bar{p}})(d_{\mathbf{a}sfp} b_{\mathbf{a}s'f'p}) e^{-4ip^0 t} + (b_{\mathbf{a}s'f'p}^+ d_{\mathbf{a}sfp}^+)(b_{\mathbf{a}sf\bar{p}}^+ d_{\mathbf{a}s'\bar{f}\bar{p}}^+) e^{4ip^0 t} + 2(b_{\mathbf{a}sfp}^+ b_{\mathbf{a}s'f'p})(d_{\mathbf{a}sf\bar{p}}^+ d_{\mathbf{a}s'\bar{f}\bar{p}}) \right), \quad (3.16)$$

where we take $V = \int d^3 x$ to be the relevant volume of integration, and a factor of 3 comes from the color matrices. It is interesting that the quark Hamiltonian has both a "free" part $H_0$ and an interaction part $H_I$ whose structure is consistent with the interpretation of hadron interactions in terms of meson exchange. It is apparent that the state $|\mathbf{x}_0\rangle$ defined above represents the ground state or 'vacuum' of the "free" theory. Due to the volume factor in $H_0$, adding any plane-wave quark or anti-quark on top of this "free" vacuum would result in an infinite increase in energy. Thus, in the context of the "free" Hamiltonian, no plane-wave quarks or anti-quarks (or colorless combinations of them) can exist. As stated earlier, only if the quark fields are confined to some finite volume will it be possible for quark states on top of the "free" vacuum to have finite energy.

The effect of the interaction Hamiltonian does not change these qualitative results. This can be seen by starting with "free" eigenstates and using a Gell-Mann and Low expansion (see [14] for a good presentation) to calculate the corresponding true energy eigenvalues. An $n$th order correction to the "free" energy will contain $n$ time integrals and $n+1$ factors of $H_I$. From (3.16), each factor of $H_I$ brings a factor of $-1$, leading to a factor of $(-1)^{n+1}$. In a Gell-mann and Low expansion, each time integral is accompanied by a factor of $-i$, leading to a factor of $(-i)^n$. Finally, after letting $e \to 0$, the $n$ time integrals produce a factor of $i^n$ in the denominator. These factors result in an overall factor of $-1$ for any full, connected contraction. As a result, the correction to the vacuum energy from $H_I$ leads to more negative energy.

Contractions that result in factors of $bb^+$ or $dd^+$ also have overall factors of $-1$. When these contractions are put into normal order to act on quark states that sit on top of the vacuum, they have a positive overall sign, adding to the infinite positive energy (relative to the vacuum) needed to create them. Contractions that result in strings of 2 or 3 destruction then creation operators can also act on "free" meson or baryon solutions. Although signs of these terms vary, their magnitude cannot exceed the positive energy of the free and $bb^+$ or $dd^+$ contraction terms. Therefore, plane-wave states involving up to three quarks or anti-quarks would require infinite energy to create and are therefore not seen. A more complete proof for all quark states is outside the scope of this paper.

By speaking in the language of the Gell-Mann and Low theorem, we have been implicitly treating $H_I$ perturbatively. From the complicated form of $H_I$, one may wonder whether any new divergent renormalization problems appear in a perturbation expansion of the quark Hamiltonian. Fortunately, these problems do not arise. This can be seen in the context of perturbation theory by noting that renormalization problems result from divergent internal loop integrals. Since every operator in $H_I$ is fixed at the same or opposite momentum, it is impossible to have internal loops for connected perturbation theory diagrams. Thus there are no divergences in the pure quark sector.

Another way to see that there are no new renormalization problems is to realize that the entire Hamiltonian $H_q$ is diagonalizable. This can be seen in the following way: First, note that $H_q$ is block diagonal in momentum – it can only connect states with the same or opposite momentum. However, for any given momentum, there are a finite number of states that can be created using the hadron operators of (3.8). At some point, adding any new hadron operator to a string of hadron operators will result in two identical quark (or antiquark) operators in a string of creation operators. Since these states vanish due to the anti-commutivity of the quark operators, the pure fermion nature of the theory puts an upper limit on the number of possible states for a given momentum. Thus the submatrix of the Hamiltonian at any given momentum is a finite dimensional matrix that can in principle be diagonalized. Since it can be diagonalized, $H_q$ cannot be divergent.

We propose that the eigenstate of the vacuum of $H_q$ can be determined in the following way: Start with $|\mathbf{x}_0\rangle$, then use the hadron creation operators of (3.8) to create every possible state $|\mathbf{x}_q^{(n)}(p)\rangle$ on top of $|\mathbf{x}_0\rangle$ for a given momentum that is flavorless and has no baryon number. Since that set of states is finite, one can create a finite-dimensional matrix $\langle \mathbf{x}_q^{(n)}(p)|H_q|\mathbf{x}_q^{(m)}(p)\rangle$. The eigenstates associated with the lowest energy eigenvalues of that matrix at each value of the momentum corresponds to the vacuum state $|0\rangle$.

**Summary and Future Work**

Using a fermionic background gluon field, we have derived new nontrivial extrema of the QCD action. These extrema involve colorless quark states that solve the equations of motion by using quantum techniques that do not have classical equivalents. In the context of these colorless quark states, we have shown that quarks and gluon fluctuations decouple in the QCD Hamiltonian, so gluon fluctuations can be ignored when calculating energy differences between two colorless quark configurations. We have also shown that the remaining quark Hamiltonian is completely diagonalizable.

Perhaps the most interesting result of this method is that the quark Hamiltonian leads to a vacuum that has infinitely negative energy. Since the Hamiltonian can be diagonalized, that quark energy can in principle be explicitly determined. This should be contrasted with standard models of the vacuum that lead to infinitely positive energies due to zero point fluctuations. Since the negative energies presented here are only possible for colorless quark states, it leads to the conclusion that since the vacuum seeks out the lowest possible energy configuration, it must involve only colorless quark states (and separately colorless gluon fluctuation states). Similarly, for hadrons to reach their lowest possible states, the quarks inside their eigenstates must also be colorless. In fact, states with colored combinations of quarks cannot be built upon the infinitely negative energy presented here, but must be built upon the infinitely positive vacuum energy of perturbative models. That being the case, colored quark states have infinitely more energy than colorless quark states, and are therefore confined.

The background gluon field that we used was given by $i \int d^4 y \, \mathbf{d}(x_0 - y_0) \partial_m \mathbf{c}^+ M \overline{T}^a \Psi$ in the Hamiltonian frame. In (2.4), we showed that up to a constant term and a term that vanishes using the quark equations of motion, the background field is also equal to $-\int d^4 y \, \mathbf{q}(x_0 - y_0) \partial_m \overline{\Psi} M \overline{T}^a \Psi$. We chose the former form for the background field due to the fact that in the Hamiltonian frame all fields have the same time coordinate. As a result, standard canonical methods can be used to calculate the Hamiltonian. It would be interesting to use the latter form instead as the definition of the background field along with appropriately generalized techniques to calculate the Hamiltonian. One of the interesting features of this second form of the background field is that it defines the field at a given time as being dependent on fields in its past (or outside of its future light cone for a general reference frame $u$). As such, this second form of the background field treats time asymmetrically, imposing an "arrow" on time.

In future papers, we will incorporate the effects of chiral symmetry breaking and quantization of quark fields inside a sphere in order to provide a more realistic picture of the vacuum and hadron masses.

**Acknowledgements**

I would like to thank Professor Suzuki of U.C. Berkeley, Professor Heinz of Ohio State University, and Doctor Lammert of Penn State University for invaluable feedback and advice in constructing this paper.

[1] M. A. Shifman, A. I. Vainshtein and V. I. Zakharov, Nucl. Phys. B147, 385 (1979); B147, 448 (1979); B147, 519 (1979).

[2] E. U. Shuryak, The QCD Vacuum, Hadrons and Superdense Matter, World Scientific, Singapre, 1988; T. Schaefer and E. V. Shuryak, Rev. Mod. Phys. 70, 323 (1998).

[3] A. Chodos, R. L. Jaffe, K. Johnson, C. B. Thorn and V. Weisskopf, Phys. Rev. D9, 3471 (1974).

[4] G. K. Savvidi, Phys. Lett. B71, 133 (1977).

[5] H. G. Nielsen and M. Ninomiya, Nucl. Phys. B156, 1 (1979).

[6] C. G. Callan, R. Dashen, D. J. Gross, Phys. Rev. D17, 2717 (1978).

[7] E. Witten, Nucl. Phys. B149, 285 (1979).

[8] A. Cabo, S. Penaranda, and R. Martinez, Mod. Phys. Lett. A10, 2413 (1995).

[9] T. Schaefer, E. Shuryak, Rev. Mod. Phys. 70, 323 (1998).

[10] T. Kugo and I. Ojima, Prog. Theor. Phys. 60, 1869 (1978); T. Kugo and I. Ojima, Prog. Theor. Phys. 61, 294 (1979); T. Kugo and I. Ojima, Prog. Theor. Phys. 61, 674 (1979); T. Kugo and I. Ojima, Prog. Theor. Phys. Suppl. 66, 1 (1979)

[11] L.F. Abbott, Nucl. Phys. B185, 189 (1981); R.B. Sohn, Nucl. Phys. B273, 468 (1986); H. Kluberg-Stern and J.B. Zuber, Phys. Rev. D12, 482 (1975).

[12] S. Weinberg, Rev. Mod. Phys. 61, 1 (1989).

[13] V. Sahni, Class. Quant. Grav. 19, 3435 (2002)

[14] "Quantum Theory of Many Particle Systems", Alexander L. Fetter and John Dirk Walecka, McGraw-Hill, 1971.